\documentclass[intlimits,twoside,a4paper]{article}

\usepackage{amsmath,amssymb}
\usepackage{graphicx}
\usepackage{wrapfig}
\usepackage[T2A]{fontenc}
\usepackage[cp1251]{inputenc}

\usepackage[eqsecnum]{cmpj2}
\issue{2012}{15}{2}{23601}
\doinumber{10.5488/CMP.15.23601}

\title[Ar--Xe mixed film on graphite]%
{On the phase behavior of mixed Ar--Xe submonolayer films on graphite}

  \author[A. Patrykiejew]{A. Patrykiejew}
  \address{Department for the Modelling of Physico-Chemical Processes, MCS University Lublin, 20031 Lublin, Poland}

\date{Received December 31, 2011, in final form February 28, 2012}
\authorcopyright{A. Patrykiejew, 2012}

\begin{document}

\newcommand{\mtau}{\mbox{\boldmath{$\tau$}}}%{\mathbf{\tau}}%{\mbox{\boldmath$\tau$}}
\newcommand{\mqu}{\mathbf{q}}
\newcommand{\mbr}{\mathbf{r}}
\newcommand{\mbR}{\mathbf{R}}

\maketitle

\begin{abstract}
Using Monte Carlo simulation methods in the canonical and grand canonical ensembles,
we discuss the melting and the formation of ordered structures of mixed Ar--Xe submonolayer films on graphite. The calculations have been performed using two- as well as three-dimensional models of the systems studied. It is demonstrated that out-of plane motion does not affect the properties of the adsorbed films as long as the total density
is not close to the monolayer completion. On the other hand, close to the monolayer completion, the promotion of particles to the second layer considerably affects the properties of mixed films. 
It has been shown that the mixture exhibits complete mixing in the liquid phase and freezes into solid phases of the structure depending upon the film composition. For submonolayer densities, the melting temperature exhibits non-monotonous changes with the film composition. In particular, the melting temperature initially increases when the xenon concentration increases up to about 20\%, then it decreases and reaches minimum for the xenon concentration of about 40\%. For still higher xenon concentrations, the melting point gradually increases to the temperature corresponding to pure xenon film. 
It has been also demonstrated that the topology of phase diagrams of mixed films is sensitive to the composition of adsorbed layers.
\pacs 68.35.Rh, 68.43.De, 68.43.Fg, 64.75.St
\end{abstract}

\section{Introduction}

It is now well known~\cite{ar01,ar02,ar03,xe02,xe03,xe04,BruchColeZar97,BruchDiehlVen07} that monolayer films of
both argon and xenon on graphite form incommensurate two-dimensional solid phases at low temperatures. However, the melting
transition of these two solid phases has a different mechanism~\cite{gr17}.
Experimental data~\cite{gr02,gr03,gr08,gr09,gr15} as well as computer simulations~\cite{grsym3,grsym10} have demonstrated
that submonolayer films of argon melt via continuous phase transition at the triple point temperature equal to
$T_\mathrm{t}\approx 49.7$~K~\cite{gr14}. At the temperatures well below the melting point, the solid-like argon
submonolayer film is rotated by about 2--3 degrees with respect to the $R30^{\mathrm{\circ}}$ axis of the commensurate $(\sqrt{3}\times\sqrt{3})R30^{\mathrm{\circ}}$ structure~\cite{grsym10,grt1,grt2,gr16}. Since the argon atoms are rather small,
the solid phase is compressed with
respect to the $(\sqrt{3}\times\sqrt{3})R30^{\mathrm{\circ}}$ commensurate structure, and the monolayer density is about 25\% higher than
the density of a perfect commensurate structure.

The melting of incommensurate submonolayer films
of xenon is of first order~\cite{xem1,xem2}, and occurs at the triple point temperature of about 100~K. However, the xenon
atoms are larger so that the incommensurate solid phase is dilated with respect to the commensurate
$(\sqrt{3}\times\sqrt{3})R30^{\mathrm{\circ}}$ structure.

The differences in the structure of low temperature phases and the phase behavior of argon and xenon submonolayer
films on graphite lead to a quite complex phase behavior of mixed films. The X-ray scattering studies of the Ar--Xe mixture
adsorbed on graphite~\cite{Bohr83,deBeau86,Ceva86} have demonstrated that the structure
of submonolayer and monolayer solid-like phases strongly depends upon the film composition.
Three different solid-like phases have been found. Apart from the compressed argon-like and dilated Xe-like incommensurate phases, being stable for sufficiently low and high xenon mole fraction, respectively, the formation
of krypton-like $(\sqrt{3}\times\sqrt{3})R30^{\mathrm{\circ}}$ commensurate structure has been found over a rather wide range of the mixture
composition. The calorimetric study of Ma et al.~\cite{Ma98} has shown that even very small amounts of xenon, about 1.5\%,
added into submonolayer argon films, lead to the disappearance of the heat capacity peak attributed to the orientational transition in the argon-like incommensurate solid phase~\cite{grsym10,grsym9}.

A vast majority of theoretical studies of mixed adsorbed layers has been based on lattice gas mo\-dels \cite{mix1,mix2,mix3,mix4,mix5,mix6}, which do not constitute a good basis for the discussion of the incommensurate-commensurate transitions
in the films of rare gases on graphite. Such models cannot properly describe the incommensurate floating solid.
However, there have also been some attempts to construct theoretical models for the commensurate-incommensurate transitions in
monolayer films of rare gas mixtures on graphite~\cite{Marti89,Villain91,Marti94}. The primary aim of the models proposed by
Marti et al.~\cite{Marti89,Marti94}, was to explain the anomaly in the phase behavior of Ar--Xe and Kr--Xe submonolayer films on graphite. Namely, experiments have demonstrated~\cite{Marti89} that less krypton than argon is needed to induce the
formation of commensurate phase. On the other hand, a rather general mean-field model of Villain and Moreira~\cite{Villain91} requires the
introduction of several approximations in order to make the resulting equations numerically tractable. Consequently,
the agreement with experimental data is rather poor. Nevertheless, these authors have derived qualitative phase diagrams
for the Ar--Xe mixture adsorbed on graphite.
The theory predicts the existence of incommensurate, Ar-like and Xe-like structures at low
temperatures, and the formation of the commensurate structure at higher temperatures over a rather limited range of the Xe mole
fraction.

In this work, we present and discuss the results of rather extensive Monte Carlo simulations of mixed Ar--Xe films on graphite.
Our main goal has been to investigate the structure of low temperature solid phases as well as to
determine the changes of the melting temperature with the mixture composition. However, we also discuss the
evolution of phase diagrams resulting from the changes in the film composition.

The paper is organized as follows. In the next section we present the model used and describe the Monte Carlo method
used to determine the properties of mixed submonolayer films. Then, in section~3 we briefly discuss the behavior of
pure Ar and Xe films. The last section~4 is devoted to the presentation of the results for the mixed Ar--Xe films.

\section{The model and Monte Carlo methods}

The interaction between adsorbate atoms is assumed to be represented by the (12,6) Lennard-Jones potential
\begin{equation}
u_{i,j}(r_{ij}) = 4\varepsilon_{i,j}\left[\left(\sigma_{i,j}/r_{ij}\right)^{12} - \left(\sigma_{i,j}/r_{ij}\right)^{6}\right],
\label{eq:01}
\end{equation}
where $r_{ij}$ is the distance between a pair of atoms and $i$ and $j$ mark the species Ar and Xe.
The values of the parameters $\varepsilon_{i,i}$ and $\sigma_{i,i}$ used in this work are given in table~\ref{tab1}.
\begin{table}[h]
\begin{center}
\caption{Lennard-Jones parameters for Ar and Xe used in this work.}\label{tab1}
\vspace{0.3cm}
\begin{tabular}{|c|c|c|}
\hline
 $i,j$   &  $\sigma_{i,j}$  & $\varepsilon_{i,j}$  \\
   &     \AA  &   K   \\    \hline
Ar,Ar  &   3.4 & 120.0  \\
Kr,Kr  &   4.1 & 221.0  \\
Ar,Kr  &   3.5 & \ \ 162.85 \\ \hline
\end{tabular}
\end{center}
\end{table}
The corresponding parameters representing the Ar--Xe interaction, also given in table~\ref{tab1}, have been obtained using
the usual Lorentz-Bertholot combining rules:
\begin{equation}
\sigma_{i,j} = \frac{1}{2}\left(\sigma_{i,i}+\sigma_{j,j}\right) \qquad \text{and} \qquad \varepsilon_{i,j}=\sqrt{\varepsilon_{i,i}\cdot
\varepsilon_{j,j}}\, .
\label{eq:02}
\end{equation}
The potential (\ref{eq:01}) has been cut at the distance $3\sigma_{i,j}$.

We are aware of some drawbacks that the assumption of the LJ potential has got, and that other authors have used different potentials
to reproduce experimental data for adsorbed films of pure rare gases on graphite~\cite{pot2}. Also, we have not taken into
account the surface mediated interactions, which are known to affect the strength of adsorbate-adsorbate interaction in the
vicinity of solid substrates~\cite{BruchColeZar97}.

The interaction of rare gas atoms with the graphite basal plane can be represented by the potential proposed by
Steele~\cite{steele1}
\begin{equation}
v_i(x,y,z)=\varepsilon_{\mathrm{gs},i}\Big[v_{0,i}(z)+\sum_{k}v_{k,i}(z)f_k(x,y)\Big], \qquad i=\text{Ar or Kr}.
\label{eq:03}
\end{equation}
In the above, the first term in the square brackets is the fluid-solid potential averaged over the entire surface,
while the second term represents the corrugation part of the fluid-solid potential.
Assuming that the interaction between an adsorbate atom and the carbon atom of the graphite substrate is also
represented by the Lennard-Jones potential, the explicit expressions for
$v_{0,i}(z)$, the Fourier components $v_{k,i}(z)$ and the functions $f_k(x,y)$ are given by the following equations:
\begin{equation}
v_{0,i}(z)=\frac{4\pi A_i^6}{a_\mathrm{s}}\sum_{n=0}^{\infty}\left[\frac{2A_i^6}{(z+n\Delta z)^{10}} -\frac{1}{(z+n\Delta z)^4}\right]\;,
\label{eq:04}
\end{equation}

\begin{equation}
v_{k,i}(z)=\frac{2\pi A_i^6}{a_\mathrm{s}}\left[\frac{A_i^6}{30}
\left(\frac{q_k}{2z}\right)^5K_5(q_kz)-2\left(\frac{q_k}{2z}\right)K_2(q_kz)
\right]
\label{eq:05}
\end{equation}
and
\begin{equation}
f_k(x,y)= \sum_l\exp[\ri\mqu_{k,l}\mtau], \qquad \mtau=(x,y)
\label{eq:05a}
\end{equation}
with the sum running over all graphite reciprocal lattice vectors of the length $q_k$.
In the above equations $A_i= \sigma_{i,\mathrm{C}}/a_1$, where $a_1 = 2.46$~{\AA} is the graphite lattice constant, i.e., the distance between
the centers of adjacent carbon hexagons, the values of $\sigma_{i,\mathrm{C}}$ and $\varepsilon_{\mathrm{gs},i}$ ($i=\text{Ar,\ Xe}$)
are given in table~\ref{tab2}, $\Delta z = 3.4$~{\AA} is the spacing between graphite planes, $a_\mathrm{s} = 5.24$~\AA$^2$ is the
area of the graphite unit cell, $K_2$ and $K_5$ are the modified Bessel functions of the second kind and of the second and fifth order respectively, and $q_k$'s are the lengths of the graphite basal plane reciprocal lattice vectors.
\begin{table}[h]
\begin{center}
\caption{The parameters describing the Ar-graphite and Xe-graphite interaction, obtained using the Lorentz-Bertholot combining rules [given by equation~(\ref{eq:02})] and assuming that $\varepsilon_\mathrm{C,C}=28$~K and $\sigma_\mathrm{C,C}=3.4$~\AA.\label{tab2}}
\vspace{0.3cm}
\begin{tabular}{|c|c|c|}
\hline
 $i$   & $\varepsilon_{\mathrm{gs},i}$ & $A_i$ \\
   &      K   &  \AA  \\ \hline
Ar & 58.00 &  3.40 \\
Kr & 78.66 &  3.75 \\
 \hline
\end{tabular}
\end{center}
\end{table}

In the case of only partially filled monolayer films and at sufficiently low temperatures, the promotion of the second
layer is likely to be negligibly small. This allows us to consider a simple strictly two-dimensional model with the external field of the form
\begin{equation}
v(x,y)= V_{\mathrm{b},i}f_1(x,y)
=-V_{\mathrm{b},i}\left\{\cos(\mqu_1\mbr)+\cos(\mqu_2\mbr)
+\cos([\mqu_1-\mqu_2]\mbr)\right\},
\label{eq:06}
\end{equation}
where the parameter $V_{\mathrm{b},i}$ ($i=\text{Ar,\ Kr}$) determines the amplitude of the corrugation potential.
The magnitudes of $V_{\mathrm{b},i}$ ($V_{\mathrm{b},\mathrm{Ar}}=0.07$ and
$V_{\mathrm{b},\mathrm{Xe}}=0.08$) have been adjusted in such a way that the results for each component are more or less consistent
with the full 3D calculations.

Simulations have been performed using the Monte Carlo method in the
canonical and grand canonical ensembles~\cite{AT87,LB00}. In the case of two-dimensional model, the rectangular simulation
cell of the size $La_1\times L\sqrt{3}a_1/2$, with $L=60$ and with the standard periodic boundary
conditions has been used. In three dimensional calculations, the simulation cell has been the rectangular
parallelepiped of the size $60a_1\times 60a_1\sqrt{3}/2\times 10a_1$, with the periodic boundary
conditions applied in the directions parallel to the substrate surface and with the reflecting hard wall located at
$z=10a_1$.

The quantities recorded included  the average potential energy, $\langle e\rangle$, the contributions to the potential
energy due to the fluid-fluid interaction, $\langle e_\mathrm{gg}\rangle$ and the contributions due to the
fluid-solid interaction for each component $\langle e_{\mathrm{gs},i}\rangle$ and the heat capacity obtained from the
fluctuations of the potential energy,
\begin{equation}
C_V= \frac{N}{kT^2}\left(\langle e^2\rangle -\langle e\rangle^2\right)  \;.
\label{eq:hcap}
\end{equation}

In order to monitor the structure of solid phases we have used radial distribution functions, $g_{ij}(r)$, for different
pairs of species $i$ and $j$, and appropriate order parameters. The formation
of hexagonally ordered phases has been monitored using the bond-orientational order parameters~\cite{katstr,SSR}
\begin{equation}
\Psi_{6,i} = \left|\frac{1}{N_{\mathrm{b},i}}\sum_{m_i}\sum_{n_i}\exp\left(\ri6\phi_{m,n}\right)\right|
\label{eq:or6i}
\end{equation}
measured separately for each adsorbate ($i=\text{Ar or Xe}$). In the above, the first sum runs over all atoms of the $i$-th component,
the second sum runs over all nearest neighbors of the same type, $\phi_{m,n}$ is the angle between the bond joining the atoms $m$ and $n$
and an arbitrary reference axis, chosen here to be the $x$-axis of the simulation cell, and $N_{\mathrm{b},i}$ is the
number of bonds between pairs of the like atoms. Also, we have calculated the total bond-orientational order parameter
\begin{equation}
\Psi_{6} = \left|\frac{1}{N_{b}}\sum_m\sum_n\exp\left(\ri6\phi_{m,n}\right)\right|,
\label{eq:or6tot}
\end{equation}
where the first sum runs over all atoms in the system and the second over all nearest neighbors.

The above defined bond-orientational order parameters make it possible to detect the hexagonally ordered structures, but are not
suitable to distinguish between the commensurate and incommensurate phases. In the commensurate phase, the atoms are localized
over the centers of carbon hexagons, and the appropriate order parameter allowing to monitor such localized structures can be
defined as~\cite{roth1999}
\begin{equation}
\Phi_i=\left|\frac{1}{6N_i}\sum_{m}\sum_{n=1}^6\exp\left(\ri\mqu_n\mbr_{m,i}\right)\right|.
\label{eq:corr}
\end{equation}
The first sum is taken over all atoms of the $i$-th component, while the second sum runs over the six reciprocal lattice vectors
 $\mqu_n$ of the graphite substrate and $\mbr_{m,i}$ is the position of the $m$-th atom of component~$i$.

The above defined order parameters have been supplemented by the corresponding susceptibilities
\begin{equation}
\chi_\mathrm{op}= \frac{L_xL_y}{kT}\left[\langle \mathrm{op}^2\rangle - \langle \mathrm{op}\rangle^2\right],
\label{eq:orsus}
\end{equation}
where `op' stands for any of the above given order parameters.

When grand canonical simulations have been carried out, we have also recorded the adsorption-desorption isotherms.

Throughout this paper, we use reduced quantities, assuming that the graphite lattice constant $a_1$
is the unit of length, and the Lennard-Jones parameter $\varepsilon_\mathrm{Ar,Ar}$ is the unit of energy. We have
decided, however, to give the
temperature in Kelvins, as it allows for easier comparison of our simulation results with experimental data. All the
densities are expressed in commensurate monolayers.

\section{The results for pure Ar and Xe films}

\begin{figure}[!t]
\centerline{\includegraphics[width=0.50\textwidth]{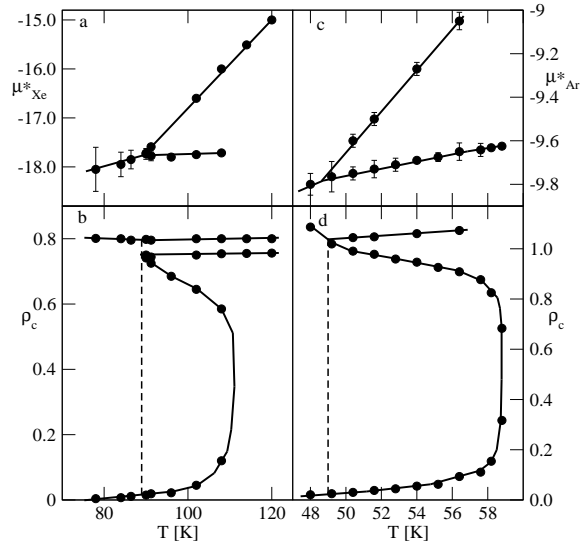}}
\caption{The phase diagrams for pure Xe (parts (a) and (b)) and pure Ar (parts (c) and (d)) monolayer films on graphite derived from
grand canonical Monte Carlo simulation. Parts (a) and (c) show the temperature-chemical potential projections, while
parts (b) and (d) show the temperature-density projections.}
\label{fig01}
\end{figure}

\begin{figure}[!b]
\centerline{\includegraphics[width=0.50\textwidth]{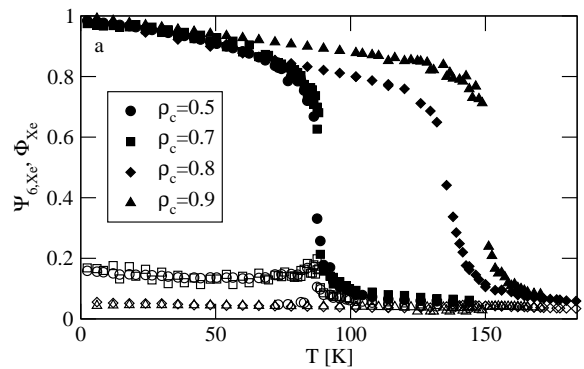}}
\caption{The temperature changes of the order parameters $\Psi_{6,\mathrm{Xe}}$ and $\Phi_\mathrm{Xe}$ for pure Xe films on graphite at different
densities (given in the figure). The results for $\rho_\mathrm{c}=0.5$ and 0.7 have been obtained using a two-dimensional model, while
those for $\rho_\mathrm{c}=0.8$ and 0.9 using a three-dimensional model. Filled and open circles mark the bond-orientational and positional  order parameters, respectively. 					}
\label{fig02}
\end{figure}

The phase diagram of xenon monolayer on graphite exhibits the vapor-liquid critical point and the triple point
(see parts (a) and (b) of figure~\ref{fig01}). The triple point temperature, equal to $T_\mathrm{tr}\approx 89$~K, agrees quite well with
some experimental data~\cite{Xetr1}, yet it is lower than the value of about 99~K stemming from other experiments~\cite{xem2,gr05}
and from the recent Monte Carlo results of Przydrozny and Kuchta~\cite{pot2}. The underestimation of the triple point temperature
is associated with our choice of the interaction potential and its parameters. Przydrozny and Kuchta applied a semi-empirical potential proposed by Aziz and Slaman~\cite{azizslaman}, while we have used a simple Lennard-Jones potential.
We should mention that in the earlier molecular dynamics simulation studies of the melting transition of xenon on graphite~\cite{far1,far2}, also based on the Lennard-Jones potential, but
with slightly different values of the parameters $\varepsilon_\mathrm{Xe,Xe}$ and $\sigma_\mathrm{Xe}$,
the triple point temperature was found to be located at $T_\mathrm{tr}\approx 0.4\varepsilon_\mathrm{Xe,Xe}/k$, while our result is
$T_\mathrm{tr}\approx 0.402\varepsilon_\mathrm{Xe,Xe}/k$.
Also, the critical point temperature $T_\mathrm{cr}\approx 109$~K  is lower than the experimental value by about 18~K~\cite{gr05}.
Nevertheless, the qualitative agreement with the available experimental data is good enough to assume that the results
for the mixed films are also qualitatively correct.

In the case of argon, the phase diagram derived from our grand canonical simulation (see parts (c) and (d) of figure~\ref{fig01}) agrees
very well with experiment. In particular, the triple point temperature, equal to $49.5\pm 0.5$~K, is practically the
same as the experimental value of 49.7~K~\cite{gr03}. Also, the critical temperature agrees very well with experiment~\cite{millot}.

The freezing of submonolayer xenon and argon films leads to the formation of incommensurate structures. The xenon incommensurate
solid attains the density of about
0.85 at the monolayer completion and is expended with respect to the commensurate $(\sqrt{3}\times\sqrt{3})R30$ structure.
Figure~\ref{fig02} shows the temperature changes of the order parameters $\Psi_{6,\mathrm{Xe}}$ and $\Phi_\mathrm{Xe}$
for xenon films of different total densities
and one sees that the order parameter $\Phi_\mathrm{Xe}$ is quite low even at very low temperatures, indicating the lack of localization of
adatoms over the minima of the graphite lattice. On the other hand, the bond-orientational order parameter $\Psi_6$ demonstrates
the formation of hexagonally ordered phase below the freezing point.

\begin{figure}[h]
\centerline{\includegraphics[width=0.45\textwidth]{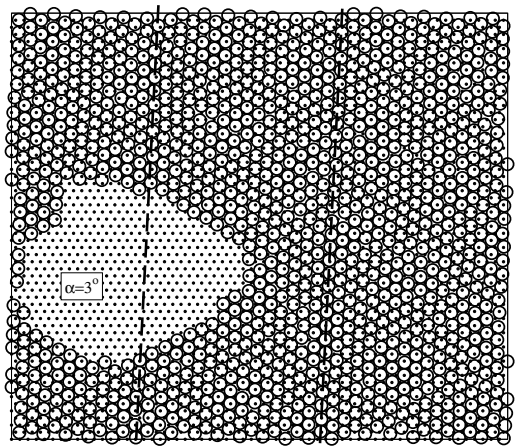}}
\caption{The snapshot of configuration for pure argon film recorded for $\rho_\mathrm{c}=1.0$ at $T=10$~K. The dashed lines show
that the film is rotated with respect to the symmetry axis of the commensurate phase by the angle $\alpha\approx 3^{\mathrm{\circ}}$. }
\label{fig03}
\end{figure}
The argon incommensurate solid also shows a well developed hexagonal symmetry, but it is contracted with respect
to the commensurate $(\sqrt{3}\times\sqrt{3})R30$ structure and attains the density of about 1.25. At sufficiently low temperatures, the film exhibits epitaxial
rotation of about 3 degrees (see figure~\ref{fig03}). Again, this result agrees very well with earlier theoretical~\cite{grt1,grt2}
and computer simulation~\cite{grsym10,grsym9} results.

The results of Monte Carlo simulation for pure argon and xenon films given above will serve as reference data for the
study of mixed films.

\section{The results for mixed films}

\begin{wrapfigure}{i}{0.45\textwidth}
\centerline{\includegraphics[width=0.4\textwidth]{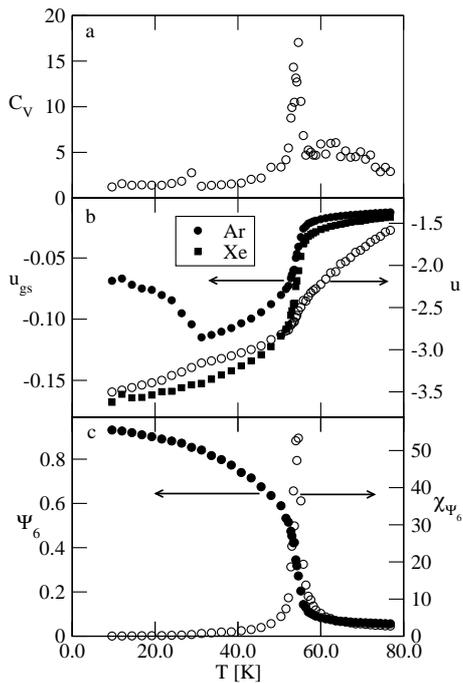}}
\caption{The temperature changes of the heat capacity (part (a)), the total potential energy and contributions to the
potential energy due to for the mixed submonolayer film of $\rho_\mathrm{c}=0.4$ and $x_\mathrm{Xe}= 0.1667$, obtained using two-dimensional model.}
\label{fig04}
\end{wrapfigure}

We begin with the presentation of canonical ensemble Monte Carlo simulation results aiming at the determination of the melting temperature and the structure of solid phases in submonolayer mixed films.
Since the solid phases (commensurate and incommensurate) exhibit hexagonal symmetry, the location of the melting point can be
estimated using the bond-orientational order parameter, $\Psi_6$, and its susceptibility, $\chi_{\Psi_6}$. Of course, one
also expects that the melting transition is manifested by sudden changes of the potential energy and the heat capacity anomalies.
Figure~\ref{fig04} gives an example of our results, obtained for the submonolayer film of the total density $\rho_\mathrm{c} = 0.4$ and the xenon mole fraction equal to $x_\mathrm{Xe}=0.1667$. Part~(a) of figure~\ref{fig04} shows the heat capacity curve and one sees a sharp peak at the melting point at $T\approx 54$~K. At the same temperature, the total potential energy $u$ and the contributions to the potential energy due to Ar-graphite and Xe-graphite interactions exhibit sudden drops (see part (b) of figure~\ref{fig04}). Finally, part~(c) of figure~\ref{fig04}, which shows the temperature changes of the bond-orientational order parameter $\Psi_6$, and its susceptibility $\chi_{\Psi_6}$ demonstrates that the melting transition is accompanied by the loss of hexagonal ordering.
It should be emphasized that a large increase of the Ar-graphite and Xe-graphite interaction energies accompanying the
freezing transition marks a sudden increase of localization of the adsorbed argon and xenon in the solid phase. Upon a
decrease of temperature, the localization of xenon gradually increases, while argon exhibits a decrease of localization
at temperatures below $T\approx 30$~K. This behavior can be attributed to the transition between the
commensurate phase, stable at $T>30$~K, and
the incommensurate phase, stable at $T<30$~K. Note that the transition is not accompanied by any changes in the
%
%\begin{figure}[h]
%\end{figure}
%
behavior of
the bond-orientational order parameter, but produces a well seen heat capacity anomaly.

\begin{figure}[!b]
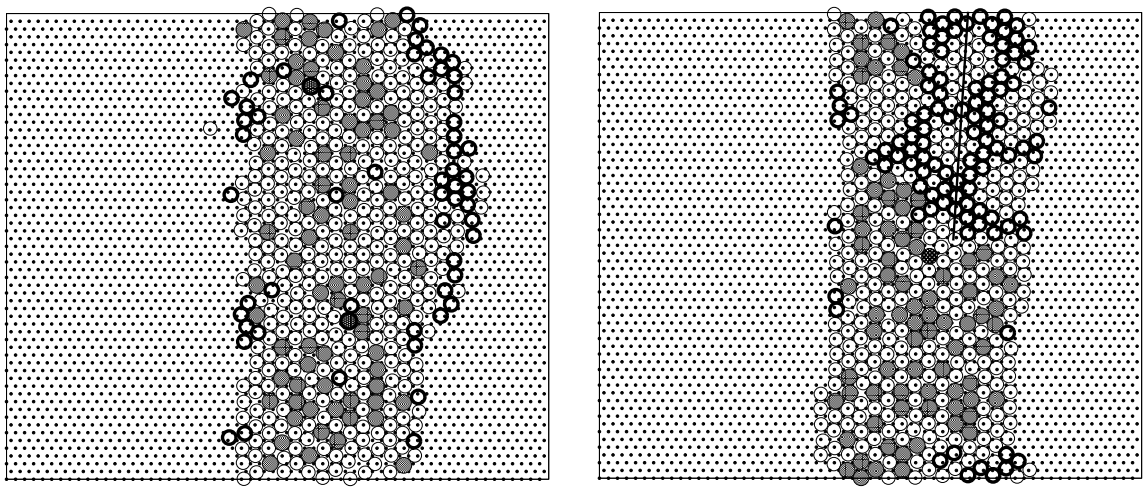

\centerline{\includegraphics[width=0.48\textwidth]{fig05a}
\hfill
\includegraphics[width=0.48\textwidth]{fig05b}}
\caption{The snapshots obtained for the film of the total density $\rho_\mathrm{c}=0.4$ and $x_\mathrm{Xe}=0.1667$ at $T=36$ (left panel) and
24~K (right panel). Black dots mark the centers of graphite cells, open circles with thin and thick lines represent argon atoms
being commensurate and incommensurate with the graphite lattice, while larger light shaded and dark shaded circles are the xenon atoms being commensurate and incommensurate with the graphite lattice. The dashed line in part (b) shows that the argon-like
incommensurate phase exhibits epitaxial rotation.}
\label{fig05}
\end{figure}

The inspection snapshots of configurations recorded during the simulation runs have shown that the commen\-surate phase is mixed,
while the incommensurate phase consists of argon only. In the case of small xenon mole fraction, as in the system considered now,
we expect to observe only a partially developed commensurate phase. Indeed, the snapshot given in figure~\ref{fig05}~(a) shows that the film
peripheries are predominantly occupied by argon atoms, which also show a rather high degree of incommensuration. At the
temperature below commensurate-incommensurate transition, we find coexisting domains of mixed commensurate and argon-like incommensurate phases (see figure~\ref{fig05}~(b)). In the snapshots given in figure~\ref{fig05}, we have assigned the atoms to commensurate and incommensurate positions using the following order parameter~\cite{grsym6}:
\begin{equation}
\phi(\mbr)=\cos(\mqu_1\mbr)+\cos(\mqu_2\mbr)+\cos([\mqu_1-\mqu_2]\mbr)\;,
\label{eq:local}
\end{equation}
and assuming that the atom is commensurate (incommensurate) when $\phi>0$
($\phi\leqslant 0$).One should note that even in a
rather small system used, consisting of only 480 atoms,
the argon-like incommensurate domain exhibits epitaxial rotation, just the same as observed for pure argon films.

\begin{figure}[ht]
\centerline{\includegraphics[width=0.55\textwidth]{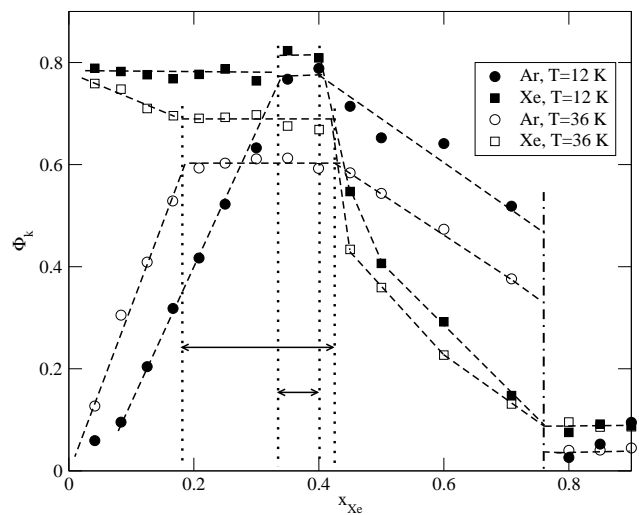}}
\caption{The order parameters $\Phi_\mathrm{Ar}$ and $\Phi_\mathrm{Xe}$ versus the xenon mole fraction in submonolayer film of the density
$\rho_\mathrm{c}=0.4$ at two different temperatures, shown in the figure. The dotted vertical lines mark the regions of $x_\mathrm{Xe}$ over
which the pure mixed commensurate phase appears, and the vertical dash-dotted line marks the xenon mole fraction at
which the domain of commensurate phase disappears.}
\label{fig06}
\end{figure}

In order to determine the locations of the commen\-surate-incommensurate transition, we have monitored the behavior
of the order parameters $\Phi_\mathrm{Ar}$ and $\Phi_\mathrm{Xe}$, defined by the equation (\ref{eq:corr}). Figure~\ref{fig06} shows the changes of these
two order parameters with the xenon mole fraction at two different temperatures in submonolayer films of the total density
$\rho_\mathrm{c}=0.4$. Quite similar results have been obtained for the films of different total densities and using two- as well as
three-dimensional models. From the observed changes of the order parameters $\Phi_\mathrm{Ar}$ and $\Phi_\mathrm{Xe}$, it follows that
the increase of the xenon concentration leads to a sequence of changes in the film structure. For small $x_\mathrm{Xe}$ we find
that xenon is highly localized, while the degree of localization of argon increases with $x_\mathrm{Xe}$. In this region,
the film consists of two coexisting phases: one being the incommensurate argon-like solid and the second being the
mixed commensurate solid. A gradual increase of the xenon mole fraction causes the commensurate domain to become larger and
the size of incommensurate domain to shrink gradually. Then, there is a region of xenon concentration over which both
adsorbates are highly localized. This corresponds to the presence of pure mixed commensurate phase and terminates
at the xenon mole fraction close to about 0.4. Then, both
order parameters gradually decrease when $x_\mathrm{Xe}$ increases up to about 0.75. In this region, the mixed commensurate phase
coexists with the demixed xenon-like incommensurate phase. Finally, for $x_\mathrm{Xe}$ exceeding about 0.75, the film
consists of xenon-like incommensurate phase with the argon atoms located at its peripheries. This has been confirmed by
the inspection of snapshots and radial distribution functions.

\begin{figure}[!t]
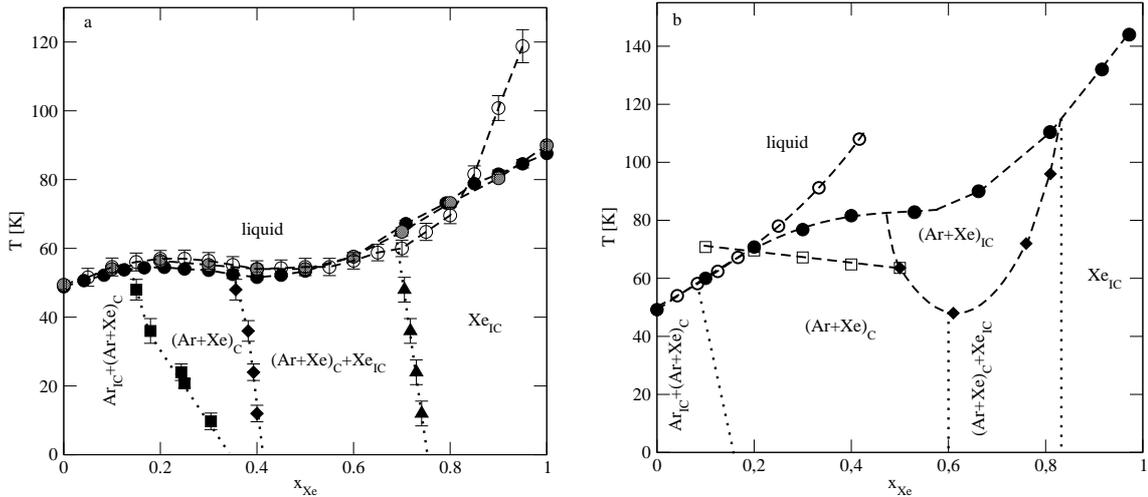

\centerline{\includegraphics[width=0.48\textwidth]{fig07a}
\hfill
\includegraphics[width=0.48\textwidth]{fig07b}}
\caption{The phase diagrams derived from the canonical ensemble simulations for submonolayer (part~(a)) and monolayer (part~(b))
mixed films of argon and xenon. In part (a), the filled, shaded and open circles show the melting temperatures for the films
of different total density equal to 0.4, 0.5 and 0.8, respectively. The results for $\rho_\mathrm{c} =0.4$ and 0.5 have been obtained using two-dimensional model, while those for $\rho_\mathrm{c}=0.8$ have been obtained using three-dimensional model. The filled squares, diamonds and triangles mark the stability regions of differently ordered solid phases. In part (b) the filled and open circles mark the
melting points obtained using three- and two-dimensional models, respectively. Open squares give the onset of the second layer
promotion and the filled diamonds show the limit of stability of the mixed incommensurate phase. Dotted lines show the
approximate locations of stability limits of different solid phases.}
\label{fig07}
\end{figure}

The central result of the canonical ensemble Monte Carlo study is given in figure~\ref{fig07}, which contains the phase diagrams showing
the locations of the melting transition and the regions of stability of different solid phases in the films of different
total densities. Part (a) of figure~\ref{fig07} gives the results for submonolayer films of different total densities, equal to 0.4, 0.667 and 0.8. The results for $\rho_\mathrm{c}=0.4$ and 0.667 have been obtained using a two-dimensional model, while those for $\rho_\mathrm{c}=0.8$ have
been obtained within a more realistic three-dimensional model. The locations of the melting point are more or less the same over a wide range of $x_\mathrm{Xe}$ between 0 and about 0.8. The independence of the melting temperature of the total density
indicates that the melting occurs at the triple point temperature. For the xenon concentration higher than 0.8 the triple point melting occurs for $\rho_\mathrm{c}=0.4$ and 0.667, but not for $\rho_\mathrm{c}=0.8$. This suggests that the density $\rho_\mathrm{c}=0.8$
is higher than the liquid density at the triple point. A rather sharp increase of the melting temperature with the xenon mole fraction, for $x_\mathrm{Xe}$ above 0.8, results from the fact that the increase of xenon concentration
brings the film closer to the monolayer completion. We should emphasize that even for $x_\mathrm{Xe}$ close to unity there is no trace of the promotion of adsorbed argon and xenon to the second layer, even at the temperatures above the
melting point. Thus, the films remain practically two-dimensional. Although we have not performed any simulation at $\rho_\mathrm{c}=0.8$ using the two-dimensional model, it can be anticipated that the results should be quite the same as those obtained with the three-dimensional model. The two-dimensional approximation is expected to
fail when the film density starts to exceed the monolayer capacity.

We have carried out the canonical ensemble calculations
assuming that $\rho_\mathrm{c}=1.0$. This density is lower than monolayer capacity of pure argon film, but it is well above
the monolayer capacity of pure xenon film.
Part (b) of figure~\ref{fig07} shows the phase diagram obtained. In this case, the
two-dimensional model works well only in the region of the xenon mole fraction not greater than about 0.2, and starts to
overestimate the stability of the solid phase for higher xenon concentrations. Three-dimensional calculations have shown
that argon is partially promoted to the second layer when the temperature becomes high enough.
The temperature at which the
second layer promotion begins depends upon the film composition and it is higher than the melting temperature only for
$x_\mathrm{Xe}$ not exceeding about 0.2 and lies below the melting temperature for higher xenon mole fractions.
It is therefore not surprising that the two-dimensional model works well only for $x_\mathrm{Xe}\leqslant 0.2$.
One of the consequences of the promotion of argon atoms to the second layer is the increase of xenon concentration
in the first layer with respect to the nominal xenon concentration in the simulation cell. The
locations of phase transitions in figure~\ref{fig07}~(b) have been plotted for the values of $x_\mathrm{Xe}$ corresponding to the actual xenon concentration in the first layer.

\begin{wrapfigure}{i}{0.5\textwidth}
\centerline{\includegraphics[width=0.40\textwidth]{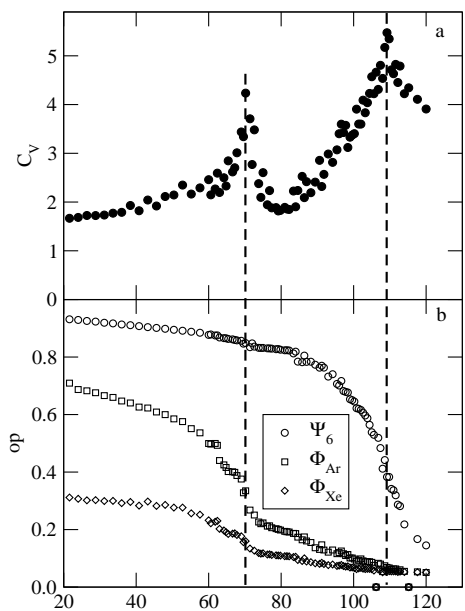}}
\caption{The temperature changes of the heat capacity (part (a)) and of different order parameters (part~(b)) obtained
for the system with $\rho_\mathrm{c}=1.0$ and $x_\mathrm{Xe}=0.7$ using three-dimensional model. The vertical dashed lines mark the freezing and commensurate-incommensurate transitions.}
\label{fig08}
\end{wrapfigure}

\begin{figure}[!b]
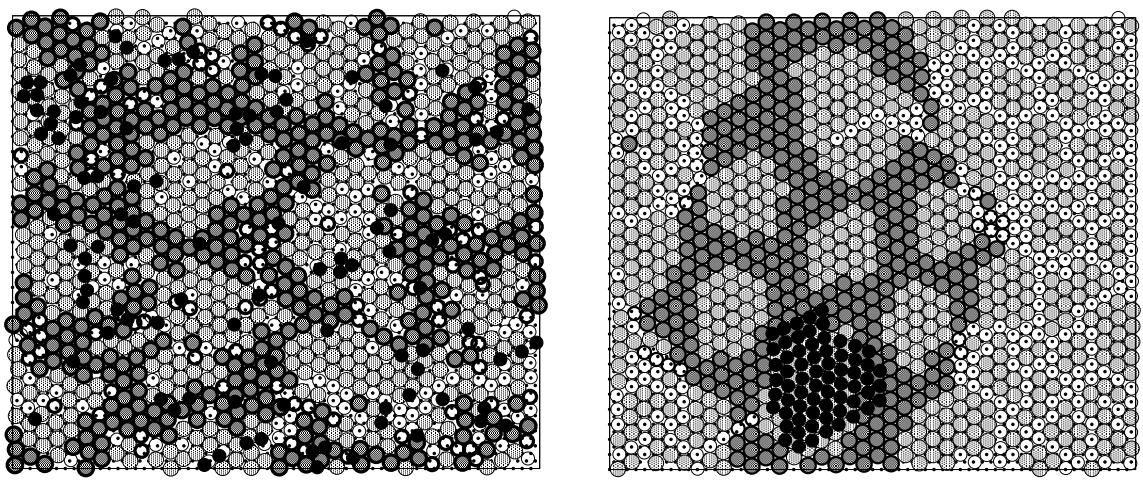

\centerline{\includegraphics[width=0.48\textwidth]{fig09a}
\hfill
\includegraphics[width=0.47\textwidth]{fig09b}}
\caption{The snapshots for the mixed film of the total density $\rho_\mathrm{c}=1.0$ and $x_\mathrm{Xe}=0.7$ at $T=72$~K (left panel) and 12~K
(right panel). Black dots mark the centers of graphite cells, open circles represent argon atoms being commensurate with the graphite lattice, while larger light shaded and dark shaded circles are the xenon atoms being commensurate and incommensurate with the graphite lattice. Filled circles stand for argon atoms located in the second layer.}
\label{fig09}
\end{figure}

One sees that the xenon mole fraction range over which the commensurate phase is stable is considerably wider
than in the previously discussed films of lower total density (cf. figure~\ref{fig07}~(a)). Moreover, we find that over a certain
range of xenon concentrations, between about 0.48 and 0.83, the mixed incommensurate phase appears at the
temperatures just below the freezing transition, whereas no trace of such a phase has been found
in submonolayer films. Upon the decrease of temperature, this phase transforms either
into the a commensurate phase, when $x_\mathrm{Xe}$ is lower than 0.6, or into the coexisting commensurate and
xenon-like incommensurate phase, when $x_\mathrm{Xe}$ is higher than 0.6. Figure~\ref{fig08} shows the changes of the heat capacity (part (a))
and the order parameters (part (b)) in the case of the film with $x_\mathrm{Xe}=0.7$. The heat capacity has two pronounced
anomalies. The first one, at $T\approx 108.5$ K, is the signature of freezing transition, accompanied by the
development of hexagonal order in the film (see the behavior of the bond-orientational order parameter $\Psi_6$ in figure~\ref{fig08}~(b)).
At the temperatures between the freezing point and the second heat capacity anomaly at $T\approx 70$~K, the solid phase is incommensurate and the order parameters $\Phi_\mathrm{Ar}$ and $\Phi_\mathrm{Xe}$ remain very small, as expected for the
incommensurate phase. The inspection
%
%\begin{figure}[ht]
%\end{figure}
%
of snapshots has shown that the incommensurate phase is mixed and the argon is partially promoted to the second layer. The second heat capacity anomaly, at $T\approx 70$ K, is due to the onset of the transition accompanied by a rather
large increase of the order parameters $\Phi_\mathrm{Ar}$ and $\Phi_\mathrm{Xe}$, indicating the increase of localization of Ar and Xe
upon the lowering of temperature. Figure~\ref{fig09} shows the snapshots recorded
at 72~K and 12~K, which demonstrate that the transition observed leads to the formation of
domains consisting of the mixed commensurate and demixed Xe-like incommensurate phases. The snapshot recorded at 12~K also demonstrates that argon is partially promoted to the second layer, and forms a compact island of a solid-like phase. It is also noteworthy that the solid-like patch of argon in the second layer is located over the demixed xenon domain rather than over the domain formed by the mixed commensurate phase. This can be readily understood by taking into account the magnitudes of Ar--Ar and Ar--Xe interaction energies, measured by the Lennard-Jones potential parameters $\varepsilon_\mathrm{Ar,Ar}$
and $\varepsilon_\mathrm{Ar,Xe}$, and of course $\varepsilon_\mathrm{Ar,Xe}$ is considerably larger than $\varepsilon_\mathrm{Ar,Ar}$ (cf. table~\ref{tab1}). When the argon atoms from the second layer are located over the pure xenon patch, each of them has three xenon atoms from the first layer as nearest neighbors. On the other hand, if the argon patch were located over the mixed commensurate patch then some nearest neighbors from the first layer would be argon atoms, and this situation is energetically less favorable.

When the xenon concentration exceeds about 0.83,
the first layer consists only of xenon, while all argon atoms are promoted to the second layer. Of course, when the
amount of xenon in the film exceeds the monolayer capacity of pure xenon film, then the excess of xenon is also located
in the second layer.

\begin{figure}[!t]
\centerline{\includegraphics[width=0.50\textwidth]{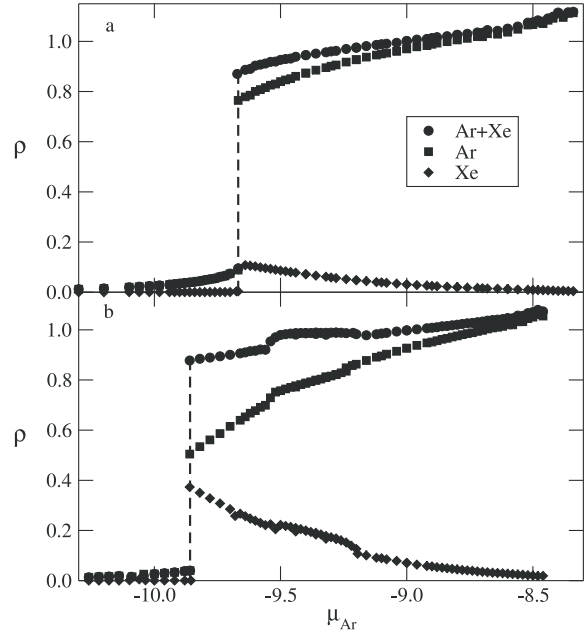}}
\caption{The examples of adsorption isotherms at $T=60$ K, obtained for $\mu^{\ast}_\mathrm{Xe}=-19.0$ (part (a)) and $-18.6$ (part (b)).
Circles show the total adsorption, while squares and diamonds denote the adsorption of argon and xenon, respectively.}
\label{fig10}
\end{figure}

We now proceed to the discussion of the changes in the phase behavior resulting from grand canonical Monte Carlo simulation.
The calculations have been performed under the condition of the fixed chemical potential of xenon, so that the xenon concentration in the film was not conserved. Along the
%
%\begin{figure}[h]
\begin{wrapfigure}{i}{0.5\textwidth}
\centerline{\includegraphics[width=0.42\textwidth]{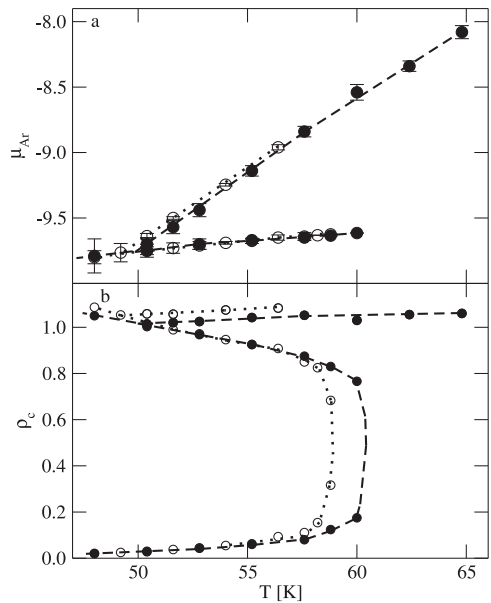}}
\caption{The phase diagram for the system with $\mu_\mathrm{Xe}=-20.0$ (filled symbols) and of pure argon film (open symbols).
Parts (a) and (b) show the temperature~-- argon chemical potential and the temperature~-- total density projections, respectively.}
\label{fig11}
\vspace{-5mm}
\end{wrapfigure}
%\end{figure}
%
adsorption isotherms obtained by changing the chemical potential
of argon, the amounts of xenon change as well. Figure~\ref{fig10} shows two examples of adsorption isotherms, both recorded at $T=60$~K, but with different values of the xenon chemical potential. It is quite evident that the gas-liquid transition is accompanied
by a sudden increase of the xenon density and that this effect is much stronger when the chemical potential of xenon
is higher. We have determined the phase diagrams for a series of systems with different values of the xenon chemical
potential, $\mu_\mathrm{Xe} = -20.0$, $-19.5$, $-19.0$ and $-18.6$.

In the case of $\mu_\mathrm{Xe} = -20.0$, the amounts of xenon in the film are very low, with $x_\mathrm{Xe}<0.01$, over the
entire range of temperatures and film densities studied. It is, therefore, not surprising that the phase diagram obtained is very similar to that of pure argon film (see figure~\ref{fig11}). In particular, the melting transition appears to be continuous and the solid phase is an incommensurate argon-like phase. However, we find that even very small amounts of xenon shift the locations of the triple and critical points towards higher temperatures.

\begin{figure}[!t]
\centerline{\includegraphics[width=0.60\textwidth]{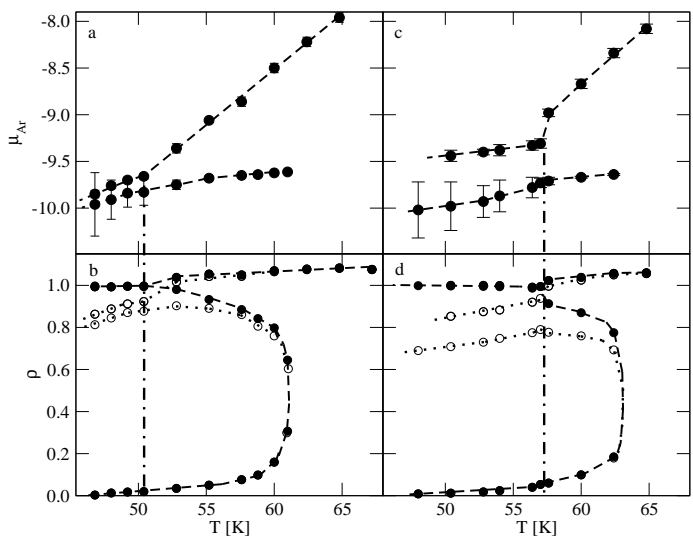}}
\caption{The phase diagrams for the systems with $\mu_\mathrm{Xe}=-19.5$ (parts (a) and (b)) and $-19.0$ (parts (c) and (d)).
Parts (a) and (c) show the temperature~-- argon chemical potential projections and parts (b) and (d) the temperature~-- total density projections, respectively. In parts (b) and (d), filled circles show the total density and open circles show the argon density.
The vertical dash-dotted lines mark the temperature above which the commensurate solid looses stability.}
\label{fig12}
\end{figure}

\begin{figure}[!b]
\centerline{\includegraphics[width=0.60\textwidth]{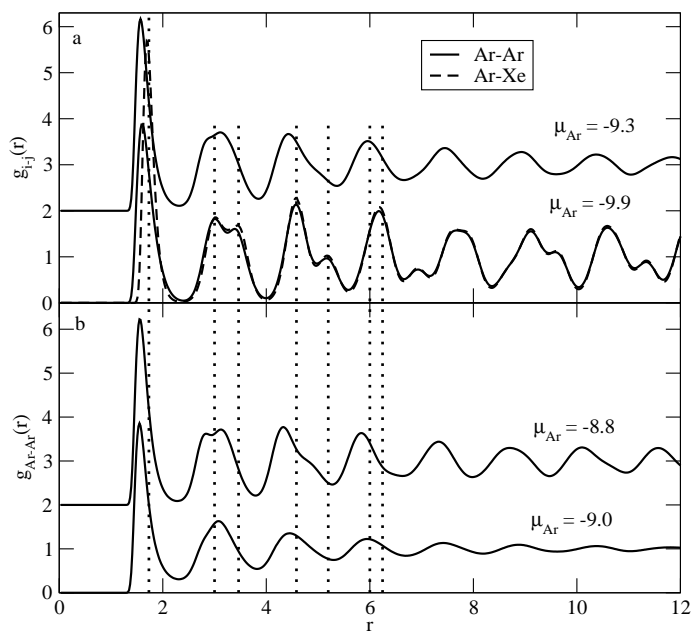}}
\caption{The argon-argon radial distribution functions for the system with $\mu_\mathrm{Xe}=-19.0$, recorded at $T=51$~K
(part (a)) and $T=57$ (part (b)) for the values of the argon chemical potential given in the figure. The vertical dotted lines
mark the locations of subsequent maxima in a perfectly ordered commensurate phase. Part (a) also shows the argon-xenon radial
distribution function obtained for $\mu_\mathrm{Ar} = -9.9$.}
\label{fig13}
\end{figure}

\newpage
An increase of $\mu_\mathrm{Xe}$ to $-19.5$ leads to some qua\-li\-ta\-ti\-ve changes in the phase diagram topology (see parts (a) and (b)
of figure~\ref{fig12}). In particular, at the temperatures below about 50.5~K, the two-dimensional gas condenses into the commensurate
krypton-like structure, which undergoes a transition into the incommensurate phase when the argon chemical potential becomes
high enough. This commensurate-incommensurate transition is continuous. One should note that the xenon concentration in the
commensurate phase is rather low ($x_\mathrm{Xe}<0.2$), and becomes still lower in the incommensurate phase.
For still higher value of $\mu_\mathrm{Xe}$ equal to $-19.0$ the phase diagram topology remains the same and only the
stability region of the commensurate phase becomes wider and extents up to $T\approx 57$ K. Also, the
commensurate-incommensurate transition occurs at higher values of the argon chemical potential. At the temperature between
the upper limit of the commensurate phase, i.e., $T\approx 57$ K, and the critical point, the gas phase condenses into the
liquid phase. The picture presented above is very well confirmed by the behavior of radial distribution functions. Part (a)
of figure~\ref{fig13} gives the argon-argon distribution functions recorded at the temperature of 51~K and for the argon chemical
potentials below and above the commensurate-incommensurate transition. It is evident that the maxima, apart from the
first one, coincide very nicely with the locations of subsequent neighbors in the commensurate phase. The stability
of commensurate phase is due to the presence of argon-xenon nearest neighbors, and the argon-xenon distribution function (also
shown in figure~\ref{fig13} as a dashed line) exhibits the first maximum very close to $\sqrt{3}a_1$, as
expected for the commensurate phase.

At the temperature of 58~K, i.e, above the upper limit of the commensurate phase stability, the argon-argon distribution function recorded at $\mu_\mathrm{Ar}=-9.0$ demonstrates the presence of a liquid phase, while at $\mu_\mathrm{Ar}=-8.8$ it corresponds to an incommensurate solid phase. The liquid is of course partially ordered due to the effects of
periodic corrugation potential.

%\begin{figure}[h]
\begin{wrapfigure}{i}{0.5\textwidth}
\centerline{\includegraphics[width=0.48\textwidth]{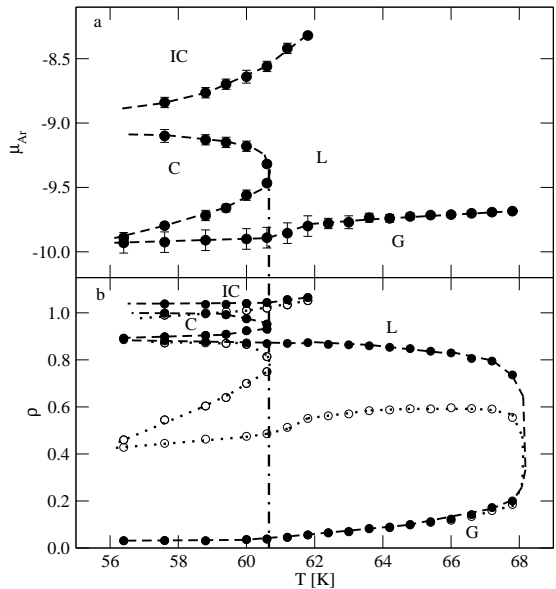}}
\caption{The phase diagram for the system with $\mu_\mathrm{Xe}=-18.6$.
Parts (a) and (b) show the temperature~-- argon chemical potential and the temperature~-- total density projections, respectively. The abbreviations G, L, C and IC stand for the gas, liquid, commensurate solid and incommensurate solid, respectively. The vertical dash-dotted line marks the temperature above which the commensurate phase does not appear.}
\label{fig14}
\end{wrapfigure}
%\end{figure}

\begin{figure}[h]
\centerline{\includegraphics[width=0.65\textwidth]{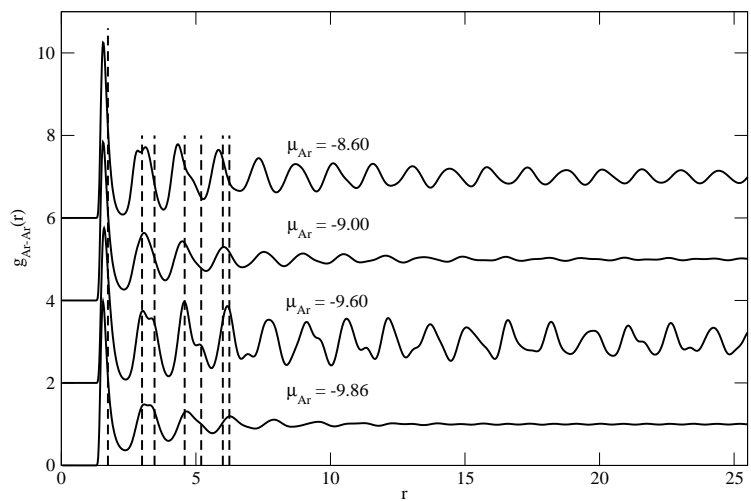}}
\caption{The argon-argon radial distribution functions for the system with $\mu^{\ast}_\mathrm{Xe}=-18.6$, recorded at $T=57$~K
and different values of $\mu^{\ast}_\mathrm{Xe}$ (shown in the figure). The vertical dotted lines
mark the locations of subsequent maxima in a perfectly ordered commensurate phase.}
\label{fig15}
\end{figure}

One should also note a gradual increase of the critical temperature resulting from the increase of xenon
concentration. The phase behavior changes when the chemical potential of xenon is increased to $-18.6$. Figure~\ref{fig14} shows that
 the gas condenses into a liquid phase of rather high xenon concentration,
ranging between $x_\mathrm{Xe}\approx 0.67$ at $T= 56$~K and $x_\mathrm{Xe}\approx 0.44$ at $T= 66$~K. When the argon
chemical potential increases, we observe the transition between the liquid and commensurate phases. This transition,
quite well illustrated by the change in the behavior of the argon-argon radial distribution function given in figure~\ref{fig15},
occurs only at the temperatures lower than 61~K. A further increase of the argon chemical potential does not lead to the transition between the commensurate and incommensurate solid phases, as in the previously
considered cases, but rather again to the liquid phase. The liquid undergoes a transition into the incommensurate solid-like
phase at still higher values of the argon chemical potential (cf. figure~\ref{fig15}). This re-entrant behavior can be understood
by taking into account that the upper limit of the film density in the commensurate phase is equal to 1.0, while the
transition into the incommensurate solid in the argon rich film and at the temperatures used occurs at the densities well
above unity. An increasing chemical potential leads to a gradual removal of xenon, so that the dense film becomes more and
more argon-like. Note that the liquid-incommensurate solid transition in pure argon film at $T=56$~K occurs at
the density of about 1.07 (cf. figure~\ref{fig01}~(b)), and at still higher densities at higher temperatures.

Concluding, we would like to present the comparison of our results with the available experimental data for submonolayer
films of the total density equal to $\rho_\mathrm{c}=0.4$.
One readily notes a qualitative agreement between Monte Carlo and experimental results. However, the present simulation
predicts a considerably narrower range of xenon concentrations corresponding to the stability region of the
commensurate phase. Unfortunately, we cannot propose any reasonable explanation for the underestimation of the
commensurate phase stability by computer simulation. One can speculate that our model based on Lennard-Jones potential and
standard mixing rules
overestimates the tendency towards demixing in submonolayer films. The commensurate phase is mixed, while the incommensurate xenon-like phase is demixed. It is also possible, however, that x-ray diffraction data overestimate the
range of xenon mole fractions corresponding to the commensurate phase. Note that within the region of coexisting
commensurate (C) and xenon-like incommensurate (IX$_\mathrm{Xe}$) phases the paches of incommensurate phase may be quite small
and hence escape detection. We recall that Villain and Moreira~\cite{Villain91} have also questioned the reliability
of experimental results given in reference~\cite{Bohr83} and suggested that the results were affected by metastability effects.

\begin{figure}[h]
\centerline{\includegraphics[width=0.65\textwidth]{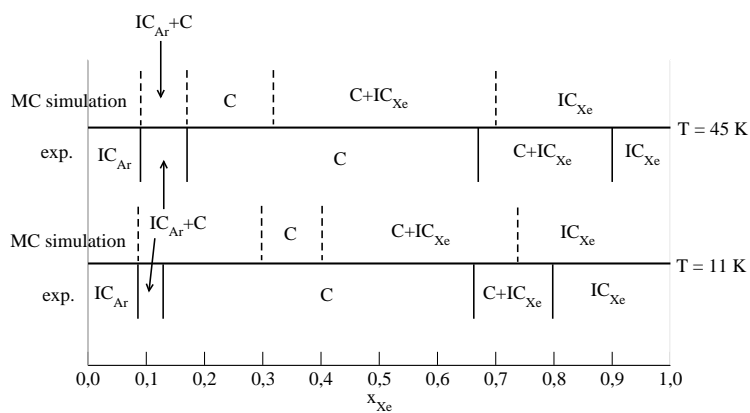}}
\caption{A comparison of the phase diagrams of Ar--Xe submonolayer films at two temperatures (given in the
figure) resulting from the present Monte Carlo simulation and from the experiment~\cite{Bohr83}.}
\label{fig16}
\end{figure}

\newpage
\section*{Acknowledgements}

This work has been supported by the Polish Ministry of Science
under the grant No. N N202 046137.

%%\section{References}
%
%{\small \topsep 0.6ex
%%\begin{verbatim}

\ukrainianpart

\title[]%
{До фазової поведінки змішаних  субмоношарових \\ плівок Ar--Xe на
графіті}

  \author[А. Патрикєєв]{А. Патрикєєв}
  \address{Відділ моделювання фізико-хімічних процесів, університет Марії Кюрі-Склодовської, \\ 20031 Люблін, Польща}

\makeukrtitle

\begin{abstract}
\tolerance=3000%
Використовуючи методи комп'ютерного моделювання методом Монте Карло у
канонічному і великому канонічному ансамблях, ми обговорюємо плавлення
і формування впорядкованих структур змішаних Ar--Xe субмоношарових
плівок на графіті. Обчислення виконуються з використанням дво- і
тривимірних модельних систем. Показано, що позаплощинний рух не
впливає на властивості адсорбованої плівки до тих пір, поки загальна
густина не стає близькою до моношарового завершення. З іншого боку,
близько до моношарового завершення, просування частинок до другого
шару значною мірою впливає на властивості змішаних плівок. 
Показано, що суміш повністю змішується в рідкій фазі і заморожується у
тверді фази зі структурою, що залежить від складу плівки. Для
субмоношарових густин, температура плавлення змінюється немонотонно зі
зміною складу плівки. Зокрема, температура плавлення спочатку зростає
з ростом концентрації ксенону близько 20\%, потім зменшується і
досягає мінімуму для концентрації ксенону близько 40\%.  Для вищих
концентрацій ксенону точка плавлення поступово зростає до температур,
що відповідають плівці чистого ксенону.  Також показано, що топологія
фазових діаграм змішаних плівок є чутливою до складу адсорбованих
шарів.
\keywords адсорбція сумішей, фазові переходи, комп'ютерне моделювання, плавлення
\end{abstract}

\end{document}